\documentclass[epj,referee]{svjour}
\input epsf.sty
\newcommand{\eqb}{\begin{equation}}
\newcommand{\eqe}{\end{equation}}
\newcommand{\dmb}{\begin{displaymath}}
\newcommand{\dme}{\end{displaymath}}
\newcommand{\pad}{\partial}
\newcommand{\ep}{\varepsilon}
\newcommand{\eab}{\begin{eqnarray}}
\newcommand{\eae}{\end{eqnarray}}
\newcommand{\ra}{\rangle}
\newcommand{\la}{\langle}
\newcommand{\e}{\mbox{e}}
\newcommand{\be}{\begin{equation}}
\newcommand{\ee}{\end{equation}}
\newcommand{\sgn}{\text{sgn}\,}

\newcommand{\text}[1]{{\rm #1}}
\begin{document}
\title{Calculation of the Regularized Vacuum Energy in 
Cavity Field Theories}
\titlerunning{Calculation of the Regularized Vacuum Energy in Cavity Field Theories}
\author{R.\ Hofmann\inst{1} \and M.\ Schumann\inst{2} \and R.D.\ Viollier\inst{2}}
\institute{Institute of Theoretical Physics,
University of T\"ubingen, 
Auf der Morgenstelle 14, 
72076~T\"ubingen,
Germany,
\and
Institute of Theoretical Physics, 
University of Cape Town,
Rondebosch 7701,
South Africa}

\mail{ralf.hofmann@uni-tuebingen.de} 
\date{\today}
\abstract{
A novel technique based on Schwinger's proper time method is applied
to the Casimir problem of the M.I.T.\ 
bag model.
Calculations of the regularized vacuum energies of massless scalar and Dirac
spinor fields confined to a static and spherical cavity are
presented in a consistent manner. While our results agree partly with previous
calculations based on asymptotic methods, the main advantage of our technique is
that the numerical errors are under control. Interpreting the bag constant as a
vacuum expectation value, we investigate potential cancellations of
boundary divergences between the canonical energy and its bag constant
counterpart in the fermionic case. 
It is found that such cancellations do not occur.
}

\maketitle

\section{Introduction}
The effects of a classical static background field on the observables of a
relativistic quantum field theory have been investigated in detail for quite some
time.  Perhaps the easiest way of distorting a free quantum field is to impose
boundary conditions on a static surface.  For the special case of an
electromagnetic field, obeying boundary conditions on two parallel, uncharged
and static plates, an attractive force was derived by Casimir as early as 1948
\cite{Cas}. 
About two decades later, great efforts were made to investigate more complicated
geometrical arrangements of the boundaries for a variety of free field
theories
and boundary conditions
\cite{Plu}.  
In particular, following the development of the
bag models of hadrons
\cite{Chodos74a,Slac,Lee}, 
there has been increased interest in
vacuum energies arising as a consequence of the boundary
conditions on a static sphere.  More
recently, such calculations have been extended to include interacting, renormalizable 
quantum field theories in the framework of perturbation theory
\cite{Bordag}. 

There are a number of approaches to calculating vacuum expectation values of
field operators in relativistic quantum field theories.  In general, one can
classify these into two categories: local or global methods
\cite{Plu}.  
The global mode
summation techniques avoid the explicit occurrence of divergences by analytical
continuation (Zeta function regularization
\cite{Nes}, 
heat kernel expansion
\cite{Bor}), 
whereas
local methods take advantage of the fact that residual local 
divergences\footnote{The free-space global divergence which renormalizes 
the cosmological constant is, in this case, already
subtracted. This corresponds to the condition that the 
vacuum expectation values vanish in the case of a free-space theory.}
manifest themselves only as nonintegrable infinities of density functions on the
boundary
\cite{Deutsch}.

The main purpose of this paper is to present a consistent calculation of the
regularized vacuum energy of massless scalar and Dirac spinor fields,
that are
confined to a static and spherical cavity, using the local stress tensor method
\cite{Deutsch,Bender,Milton1}. The coefficients of the leading boundary 
divergences of the cavity energy have already been 
found elsewhere \cite{Bender,Milton,Milton2} based on 
asymptotic expansions. However, as one of these authors points out \cite{Milton2}, 
the validity of this method can be questioned. It is therefore useful 
to check these 
results using an alternative numerical procedure. 
Moreover, the  method developed below gives also 
insights into the order of magnitude of the 
coefficients of subleading divergences and finite parts. 
In principle, these can be calculated to any
given accuracy, but in practice constraints on the computational effort can cause 
the errors to remain quite large.

In the framework of the M.I.T.\ bag  model
\cite{Chodos74a}, 
there is for each field (scalar, fermionic and vector) a linear
boundary condition that determines the eigenmodes and their energies.  There is
also a nonlinear boundary condition that guarantees the balance of the pressures
at the boundary.  The latter is usually taken into account in an average
fashion,
by minimizing the total energy of the bag  with respect to the bag radius. 
Comparing the result of this minimization with the experimental data, a
phenomenological value for the bag constant is obtained.  However, the nonlinear
boundary conditions of the model can also be used to interpret 
the bag constant as a
vacuum expectation value which makes this additional and rather artificial
minimization condition redundant
\cite{Milton}.

This paper is organized as follows:  In section II, we illustrate the local
stress tensor method in the case of the straight-forward example of scalar and
Dirac spinor fields satisfying linear boundary conditions on a plane.  
Well-known results are reproduced using the momentum-space representations of 
the free-space propagators, the reflection method 
\cite{Hansson}, 
and Schwinger's spectral $z$-forms
\cite{Stoddart,Schreiber92a,Schreiber92b,Stoddart91,Stoddart92,Page93,Cuthbert93,Viollier96}.
Section III contains a derivation of the
expressions for the 00-component of the canonical energy-momentum tensor,
integrated over the angles, for 
Klein-Gordon and Dirac fields satisfying the linear boundary
conditions of the M.I.T.\ bag model on a static, spherical surface.  
Using the cavity mode representation of the propagators
\cite{Stoddart,Buser,OConnor}, 
the numerical evaluation is done by summing over the cavity mode
quantum numbers first and integrating the resulting spectral $z$-form
subsequently.  An integration over a regularization volume yields the canonical
regularized energy.  Thereby an expansion in terms of the regularization
parameter is derived from the expansion of the density into a Laurent type series 
\cite{Deutsch}. In section IV, the expression for the fermionic bag constant is derived and
evaluated in analogy to section III.  In the last section, we summarize our 
results and 
compare them with those available in the literature, and subsequently,
 we discuss problems and speculate on their possible solutions.

\section{Vacuum energy in half-space field theories}
In this section a method for calculating the vacuum expectation value of the
canonical energy density 
$\theta^{00}$  is illustrated
for half-space field theories.  Let us
consider a massless scalar field subject to Dirichlet (D) and Neumann (N)
boundary conditions, and a massless Dirac field satisfying the linear boundary
condition of the M.I.T.\ bag model (MIT,q)
\cite{Chodos74a}
\eab
{\text D}:&&\  \phi (x)|_{x\in S}=0 \label{cond1}\\ 
{\text N}:&&\  \pad_{\mu}n^\mu\phi (x)|_{x\in S}=0\\ 
{\text {MIT,q}}:&&\  (in_{\mu}\gamma^{\mu}-1)\psi (x)|_{x\in S}=0\ ,
\label{cond3}\\  
&&n:=(0,0,0,1),\ \ S:=\{x|\ x^3=0\}.\nonumber
\eae
Using the reflection method
\cite{Hansson}, the half-space propagators read
\eab
\Delta_{\text D} (x,y)&=&\Delta^0 (x-y)-\Delta^0 (x-y^T)\\ 
\Delta_{\text N} (x,y)&=&\Delta^0 (x-y)+\Delta^0 (x-y^T)\\
S_{\text {MIT,q}} (x,y)&=&S^0(x-y)-i\gamma^3S^0(x-y^T)\ ,\\ 
y^T&=&(y^0,y^1,y^2,-y^3)\ ,\nonumber
\eae
where $S^0$ and $\Delta^0$ stand for the free-space propagators.  The 
structure of the canonical 
energy-momentum tensor $\theta^{\mu\nu}$ (see for example \cite{Tarrach}) 
requires that, in order to obtain the (diverging) vacuum energy densities,
some bilinear 
operators must be applied to the corresponding propagators. 
For a massless Klein-Gordon field, we obtain
\eqb
\label{OS}
\langle\theta^{00}(x)\rangle = \lim_{y\to x}\pad_{x^0}\pad_{y^0}\ \Delta(x,y)\ ,
\eqe
whereas, in the case of a massless Dirac field, the canonical 
vacuum energy density reads
\eqb
\label{OD}
\langle\theta^{00}(x)\rangle = 
\lim_{y\to x}\frac{i}{2} 
\mbox{Tr}\left[(\pad_{x^0}-\pad_{y^0})\gamma^0\right]\ S(x,y)\  ,
\eqe
where $\Delta(x,y)$ and $S(x,y)$ denote the field propagators, 
respectively.

Let us now give a general illustration of the $z$-form method for the calculation 
of the vacuum expectation
value of local field bilinears. 
As an example, we concentrate on the expectation value of
a Lorentz scalar in a scalar field theory,
characterized by the bilinear $\cal B$.\footnote{The Dirac case can be 
treated in straightforward analogy.} This scalar 
quantity can be expressed in terms of the identity
\eqb
\la {\cal B}(\phi(x),\phi(x))\ra=\lim_{y\to x} 
{\cal B}_{x,y}\ \Delta^0(x-y)\ .
\eqe
Here $\Delta^0(x-y)$ stands for the free-space Feynman propagator, and 
${\cal B}_{x,y}$ denotes the 
point splitted version of ${\cal B}$.\footnote{
A  contribution of a product of field derivatives
$\partial_\mu\phi(x)\,\partial_\nu\phi(x)$ to ${\cal B}$, e.g., implies a
contribution of $\partial_{x^\mu}\partial_{y^\nu}$ to ${\cal B}_{x,y}$.} 
In momentum space, 
the propagator is given by 
\eqb
\Delta^0(p)\propto\frac{1}{p^2}\, ,
\eqe
and the application of ${\cal B}$ results in 
\eqb
{\cal B}_{x,y}\ \frac{\e^{-ip(x-y)}}{p^2}\propto\ \frac{b(p^2)}{p^2} \ \e^{-ip(x-y)}\ , 
\eqe
where $b$ is a polynomial in $p^2$ (in a scale invariant theory $b$ is just a power in $p^2$
corresponding to the mass dimension of ${\cal B}(\phi(x),\phi(x))$). 
We now take the limit $y\to x$ and rotate to Euclidean momentum space 
\dmb
p_0\ \to\ ip_0\ .
\dme
The (Euclidean) denominator can then be elevated into an exponent
\eqb
\label{shift}
\frac{1}{p^2}\ \to \int_{0}^{\infty}dz\ \e^{-zp^2}\ ,
\eqe
and the remaining task is to integrate over 
the (Euclidean) momentum which involves (modified) Gaussian integrals
\eab
\label{gauss}
\lefteqn{
\int\frac{dp_{\mu}}{(2\pi)}\ \{1,\ (p_{\mu})^2,\ \dots\}\ 
\e^{-z(p_{\mu})^2} = }
\hspace{1.5cm}\nonumber\\
&&\left\{1,\frac{1}{2z},\ \dots\right\} 
\frac{1}{\sqrt{4\pi z}}\ , 
\quad \mu = 0, \dots, 3\, . 
\eae
\noindent
Calculating $\la\theta^{00}\ra$ for a massless scalar field, 
distorted by a Dirichlet plate at $x^3=0$, is 
now straightforward. The full propagator for this problem reads
\eab
\label{dplaProp}
\Delta(x,y)&=&\Delta^0(x-y)-\Delta^0(x-y^T),\\ 
 y^T&=&(y^0,y^1,y^2,-y^3)\ . \nonumber
\eae
Renormalizing the cosmological constant, by applying the bilinear operator
of Eq.\ (\ref{OS}) to the 
{\em difference} of the half- and free-space propagators, yields finite 
results except at the boundary. In the case of Dirichlet boundary conditions, we obtain
\eab
\label{zformd}
\left\la\theta^{00}(x^3)\right\ra_{D}&=&-\frac{i}{(2\pi)^4}
\int d^4p\ \frac{(p^0)^2}{p^2+i0}\ \e^{-2ix^3p^3}\nonumber\\ 
&=&\frac{1}{(2\pi)^4}
\int_e d^4p\ \frac{(p^0)^2}{p^2}\ \e^{-2ix^3p^3}\nonumber\\ 
&=&\frac{1}{32\,\pi^2}
\int_0^{\infty}dz\ \e^{-(x^3)^2/z}\ z^{-3}\nonumber\\ 
&=&\frac{1}{32\, \pi^2\,(x^3)^4}\, ,
\eae
while for Neumann boundary conditions we have
\eqb
\label{zformn}
\left\la\theta^{00}(x^3)\right\ra_{N}=-\frac{1}{32\ \pi^2(x^3)^4}\ .
\eqe
In the case of a massless Dirac field, subject 
to the 
linear boundary condition of the M.I.T.\ bag model for the quark field (see
Eqs.\ (\ref{cond1})--(\ref{cond3}),
we obtain
\eqb
\left\la\theta^{00}(x^3)\right\ra_{\text{MIT,q}}\equiv 0\ .
\eqe
This result is a consequence of 
the tracelessness of the energy-momentum tensor for this field \cite{Deutsch}.

\section{Vacuum energy in cavity field theories}
In this section, we shall calculate the canonical energy density of the vacua of
massless scalar and Dirac fields, confined to a static and spherical
cavity with radius $R$.  After a brief description of the methods, the
regularized vacuum energy is calculated for each field.

\subsection{Massless scalar fields}   
In order to perform the free-space subtraction in the spherical symmetric case,
we need to express the free scalar propagator in terms of angular
momentum eigenstates using the Rayleigh expansion for plane
waves  
\eab
\Delta^0(x-y)&=&-\frac{i}{(2\pi)^4}\int d^4p\ 
\frac{\e^{-ip(x-y)}}{p^2+i0}\nonumber\\ 
&=&-(4\pi)^2 \frac{i}{(2\pi)^4}\int d^4p\ 
\frac{1}{p^2+i0}\nonumber\\ 
& &{}\times\sum_{{l,l^\prime}\atop {m, m^\prime}}i^l(-i)^{l^\prime} 
j_l(pr)j_{l^\prime}(pr')\ 
 Y_{l,m}(\hat{r})\ \\[-0.4cm]
&&\phantom{{}\times\sum_{{l,l^\prime}\atop {m, m^\prime}}i^l} 
{} \times
Y^*_{l,m}(\hat{p})\ 
Y^*_{l^\prime,m^\prime}(\hat{r}')\ Y_{l^\prime,m^\prime}(\hat{p}) \ .
\nonumber
\eae
Applying the bilinear operator of Eq.\  (\ref{OS}) to this representation of the propagator, 
and performing in turn the angular integration, the summation 
over $m$ and a Euclidean rotation 
yields, after introducing  the $z$-integration 
according to Eq.\ (\ref{shift}), 
\eab
\left\la\theta^{00}(r)\right\ra &=& \frac{1}{2\pi^{5/2}}
\int_0^{\infty}\!\! dz\int_{0}^{\infty}\!\!dk \ k^2 \sum_{l}(2l+1)
\nonumber\\ 
&& \hspace{1.5cm} {}\times (j_l(kr))^2\, \frac{\e^{-zk^2}}{z^{3/2}}\ .
\eae
For the cavity part, we obtain the vacuum expectation value of 
the canonical energy density by 
applying the bilinear operator of Eq.\ (\ref{OS}) to the cavity mode representation of the propagator 
(see Appendix \ref{appscalar}).
A $z$-integration is introduced, which originates from a shift of the (Euclidean) momentum 
squared denominator of the propagator 
into an exponential, as discussed in the preceding section. Here the meaning of
the term momentum differs somewhat from that of free space due to the boundary
conditions, i.e.\ the analogue to the  expression
\eqb
\label{pfree}
p^2=(p^0)^2-(\vec{p})^2
\eqe
in free space, is in the cavity
\eqb
\label{pcav}
(p_{n,l})^2=\omega^2-(\ep_{n,l})^2\ ,
\eqe
where
$\omega$
denotes the arbitrary (off-shell) energy, and $\ep_{n,l}$ stands for the
energy of the mode, labelled by the radial quantum number $n$ and angular
momentum quantum number $l$.

The result of the free-space subtracted canonical vacuum energy 
density in the cavity is then
\eab
\label{angints}
\lefteqn{
\left\la\tilde{\theta}^{00}(r)\right\ra_{\text {D,N}}
:=4\pi\ \left\la
\theta^{00}(r)\right\ra_{\text{D,N}}=
}\nonumber\\ 
&=&-\frac{1}{4\ \pi^{1/2}} 
\int_0^{\infty}dz \ \frac{1}{z^{3/2}}\sum_{l}
\nonumber\\
&&\quad {}\times 
\left\{
\sum_{n,\mu}\ \int d\Omega\ [
a^{D,N}_{n,l,\mu}(\vec x) a^{*\ \text{D,N}}_{n,l,\mu}(\vec x)] \ 
\e^{-z \ep_{n,l}^2}\right.
\nonumber\\ 
& &\quad\quad \left.{}-\frac{2(2l+1)}{\pi}
\int_0^{\infty}dk\ k^2 (j_l(kr))^2\ \e^{-zk^2}\right\}\nonumber\\  
&=&-\frac{1}{4\ \pi^{1/2}} 
\int_0^{\infty}dz \ \frac{1}{z^{3/2}}\sum_{l}(2l+1)\nonumber \\
&&\quad {}\times \left\{
\sum_{n}\ \frac{1}{R^3}\ {\cal N}^{2\ \text{D,N}}_{n,l}\ (j_l(|\ep_{n,l}|r))^2\ 
\e^{-z \ep_{n,l}^2}\right.
\nonumber\\ 
& &\quad\quad  \left.{}-\frac{2}{\pi}
\int_0^{\infty}dk\ k^2 (j_l(kr))^2\ \e^{-zk^2}\right\}\ . 
\eae
Here the $a^{D,N}_{n,l,\mu}(\vec{x})$ denote the scalar cavity modes for either 
Dirichlet or Neumann
boundary  conditions, and the ${\cal N}^{\text{D,N}}_{n,l}$ 
stand for their 
normalization constants \cite{Stoddart,OConnor}, respectively, 
as explained in   Appendix \ref{appscalar}.  Using the plane-wave
representation of the free-space propagator implies a $z^{-3}$ divergence in
the free-space part of Eq.\ (\ref{angints}). 
Since finite sums over linearly
independent functions cannot change the divergence structure  common 
to all of these
terms, a numerical evaluation of 
$\la \tilde{\theta}^{00}(r)\ra_{\text{D,N}}$ 
would diverge.

The calculation reveals a substantial difference between the Dirichlet and the
Neumann case.  The Dirichlet boundary condition yields a $z$-form that is
integrable at $z=0$, whereas the Neumann boundary condition leads to a
nonintegrable $z^{-3/2}$ divergence.  We are able to show that this 
divergence is a
global one, i.e.\ independent on $r':=r/R$. It therefore resembles another
volume divergence (as does the free-space global divergence), and hence it can
be omitted through a renormalization of the cosmological  
constant.

\subsection{Massless Dirac fields}
Using Eq.\ (\ref{OD}), the mode summation representing the cavity Dirac
propagator
\cite{Stoddart,OConnor}, 
the spherical representation of the free-space propagator, and
introducing the Schwin\-ger $z$-integral, 
we obtain, after an angular  integration,
\eab
\label{angint}
\lefteqn{
\left\la\tilde{\theta}^{00}(r)\right\ra_{\text {MIT,q}}:=
4\pi\ \left\la
\theta^{00}(r)\right\ra_{\text {MIT,q}}}\hspace{0.5cm}\nonumber\\ 
&=&\frac{1}{2\ \pi^{1/2}} 
\int_0^{\infty}\!\!\! dz \, \frac{1}{z^{3/2}}
\nonumber\\
&&\ \left\{\sum_{\kappa}
\frac{1}{2}\sum_{n,\mu}\int {d\Omega}\ [q^\dagger_{n,\kappa,\mu}
(\vec{x})\ q(\vec{x})_{n,\kappa,\mu}]\ 
\e^{-z\ep_{n,\kappa}^2}\ \right.
\nonumber\\ 
& &\ \left.{}-\sum_{l}\ \frac{4}{\pi}\  (2l+1)
\int_0^{\infty}dk\ k^2 (j_l(kr))^2\ \e^{-zk^2}\right\}\nonumber\\ 
&=&\frac{1}{2\ \pi^{1/2}} 
\int_0^{\infty}\!\!\!\!dz \, \frac{1}{z^{3/2}}
\nonumber\\
&&\ \left\{\sum_{\kappa}
\frac{1}{2}\sum_{n}\frac{1}{R^3}\, {\cal N}^2_{n,\kappa}\, (2J+1)
 \left((j_{l}(|\ep_{n,\kappa}|r))^2\right.\right.
\nonumber\\
&& \qquad\qquad\qquad\qquad\qquad\left. {}+
(j_{\bar{l}}(|\ep_{n,\kappa}|r))^2\right)
\e^{-z \ep_{n,\kappa}^2}
\nonumber\\ 
& &\ \left.{}-\sum_{l}\ \frac{4}{\pi}\  (2l+1)
\int_0^{\infty}\!\!\! dk\, k^2 (j_l(kr))^2\, \e^{-zk^2}\right\} ,
\eae
where
\eab
\label{def}
J&:=&|\kappa|-\frac{1}{2}\ ,\\ 
l&:=&|J|+\frac{1}{2}\ \text{sgn}\ \kappa\ ,\\
\bar{l}&=&l-\mbox{sgn}\ \kappa\ .
\eae
Here, $q_{n,\kappa,\mu}(\vec{x})$ 
denotes the Dirac cavity mode, labelled with the radial quantum number
$n$, the Dirac quantum number $\kappa$, and the angular momentum projection
$\mu$, as discussed in  Appendix \ref{appfermi}.

\subsection{Numerical evaluation of \boldmath $\la\tilde{\theta}^{00}\ra$}
Equations (\ref{angints}) and (\ref{angint})
are suitable for a numerical evaluation.  The $k$
integration is performed in the range from zero to $\ep_{\text{max}}$, the
maximal energy eigenvalue used in the sum over cavity modes.  In our computations it
is typically of the size  $200/R$.

The calculation of $\la\tilde{\theta}^{00}\ra$ is done in two steps.  
At first, we compute the 
$z$-form for the corresponding field and linear boundary condition at a number
$M$ of points $r'$ ($M\approx 500$ and $r':= r/R$).  
It thereby proves convenient to
make the variable substitution $z=y^2$ resulting in a pure $y^{-2}$ divergence
in the $y$-form.  In a second step, we integrate the regular part of the $y$-form
and hence determine the energy density as a function of the position $r'$.

\begin{figure}
\hfil\epsfbox{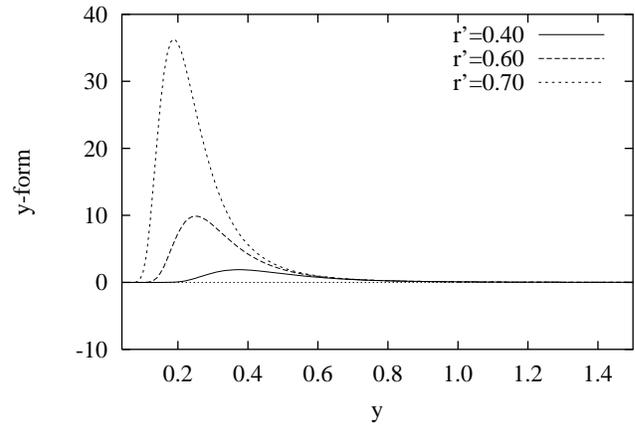}\hfil 
\caption{\label{Di} $y$-forms for $\la\theta^{00}\ra$ 
at three different locations
$r'=r/R$ for a massless scalar field 
fulfilling Dirichlet boundary
conditions on a static sphere with radius $R$.} 
\end{figure}

\begin{figure}
\hfil\epsfbox{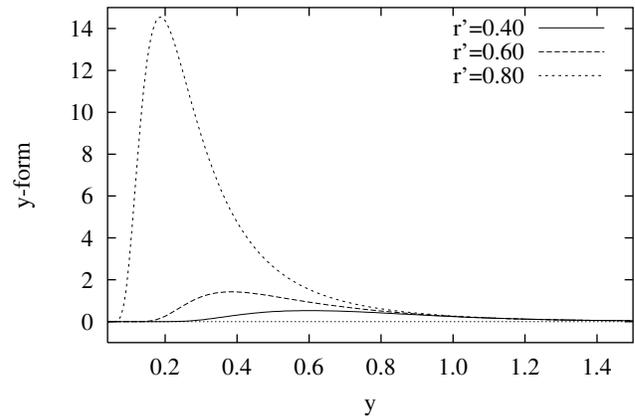}\hfil
\caption{\label{q}Massless fermion field $y$-forms 
at three different locations
$r'=r/R$. The field modes fulfill the linear boundary condition
 of the M.I.T.\ bag model 
on a static sphere with radius $R$.}
\end{figure}

\begin{figure}
\hfil\epsfbox{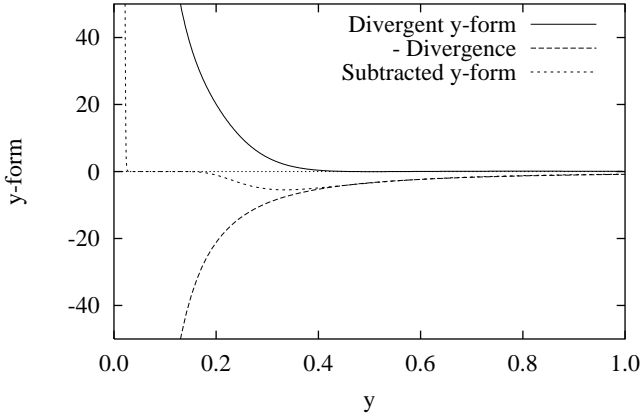}\hfil
\caption{\label{N}Unsubtracted and subtracted $y$-forms for $\la\theta^{00}\ra$ 
at $r'=0.5$ for a massless scalar field fulfilling Neumann boundary
conditions on a static sphere with radius $R$. The subtracted divergence is of the form $m/y^2$, where 
$m=0.8463$.}
\end{figure}

In Figs.\ \ref{Di} and \ref{q}, the $y$-forms for $\la\tilde{\theta}^{00}\ra$ 
are displayed for the scalar Dirichlet and the Dirac case, respectively. 
It is shown numerically that the region
of small $y$, where the form is practically zero, moves to the left with
increasing maximal energy $\ep_{\text{max}}$, whereas points to the right of
this region remain unchanged.  
The $y$-form for $\la\tilde{\theta}^{00}\ra$ in the scalar Neumann case contains 
a  (under an
increase of $\ep_{\text{max}}$) stable and global, i.e.\ independent of $r'$,
$y^{-2}$-divergence (see Fig.~\ref{N}).  The subtraction of this 
divergence amounts to renormalizing  the cosmological constant and
results in a $y$-form shown in Fig.~\ref{N}.

\begin{figure}
\hfil\epsfbox{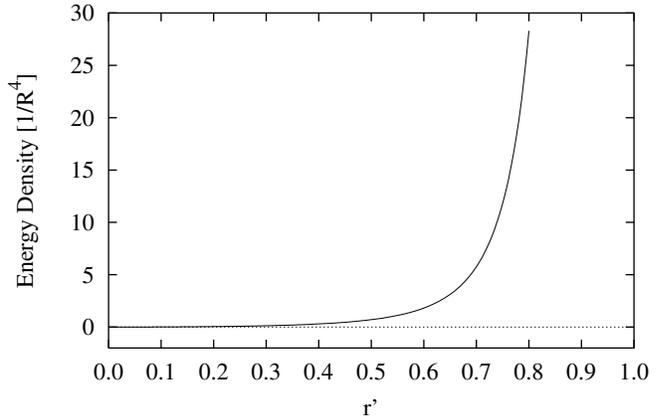}\hfil
\caption{\label{Dir}Vacuum energy density for a massless scalar field 
fulfilling Dirichlet boundary
conditions on a static sphere with radius $R$.}
\end{figure}

The error $f(r',\ep_{\text{max}})$ of 
$\la\tilde{\theta}^{00}\ra$ 
due to the truncation 
of the sum over cavity energies in Eqs.\ (\ref{angints}) and (\ref{angint})
 can be determined by varying the cut-off energy 
$\ep_{\text{{max}}}$. 
We can safely estimate the upper bound for $f(r',\ep_{\text{max}})$ 
at $10^{-6}$,
valid for all $r'$ and all fields  
and boundary conditions under consideration.
As an example, Fig.~\ref{Dir} shows 
$\la\tilde{\theta}^{00}(r')\ra$ 
for a Dirichlet scalar field.

Following Deutsch and Candelas \cite{Deutsch}, we may expand
 $\la\tilde{\theta}^{00}\ra$ into a Laurent type series around $r'=1$  
\eab
\la\tilde{\theta}^{00}\ra(\delta')&=&\frac{1}{R^4}\ 
\bigg[c_{-4}\ (\delta')^{-4}+c_{-3}\ (\delta')^{-3}+
\dots \nonumber\\
&&
\ {}+c_{-1}\ (\delta')^{-1}+c_0+\dots+c_{N}\ (r')^N\bigg] ,
\label{Las}
\eae
\dmb
N>0\ , \qquad \delta':=1-r'\ .  
\dme  
To extract the coefficients of the negative powers in Eq.\ (\ref{Las}), 
the function 
\dmb
E^L(\delta'):=c_{-4}\ (\delta')^{-4}+\dots+c_{-1}\ (\delta')^{-1}
\dme
is fitted to the calculated curve in the interval 
\dmb
\delta'_{\text{min}}\le\delta'\le\delta'_{\text{max}}\ , 
\dme
where $\delta'_{\text{min}}$ and 
$\delta'_{\text{max}}$ are 
close to zero. 
For the determination of the positive power 
coefficients in Eq.\ (\ref{Las}),  
we fit the polynomial
\dmb
E^T(r'):=a_0+a_1\ r'+\dots+(r')^N
\dme
to the calculated curve within the interval 
\dmb
0\le r'\le r'_{\text{max}}:=
1-\delta'_{\text{max}} \ .
\dme
A value of $N=9$ is used to obtain negligible fitting errors. Fitting
 $E^L(\delta'(r'))$ to the polynomial
\dmb
b_0+b_1\ r'+\cdots+b_N\ (r')^N
\dme
within the above interval, the expression for 
$\la\tilde{\theta}^{00}\ra$ reads
\eab
\label{fittheta}
\la\tilde{\theta}^{00}(r')\ra &=&E^L(\delta'(r'))+
\Delta E(r'), \nonumber\\ 
\Delta E(r')&:=&c_0+c_1\ r'+\cdots+c_N\ (r')^N\ , \\
c_i&:=& a_i-b_i, \qquad (0\le i\le N)\ . \nonumber
\eae
The numerical errors of the coefficients $c_{-4},\dots,c_{-1}$ can be estimated by varying 
the interval 
\dmb
\delta'_{\text{min}}\le\delta'\le
\delta'_{\text{max}}\ .  
\dme
In our computations we vary $\delta'_{\text{min}}$ from 0.05 to 0.15 
and $\delta'_{\text{max}}$ from 0.1 to 0.2 to obtain errors of about 1\%, 10\%
 and up to 130\% for the coefficients of the leading, next to leading, 
and the weakest divergences, respectively.

In the literature, there have been  several attempts to extract boundary divergences 
of $\la\tilde{\theta}^{00}\ra$ or the canonical energy $E(\ep')$ using the
asymptotic properties of analytic functions 
\cite{Deutsch,Bender,Milton,Rav}. 
The drawback of these expansions lies in the 
fact that one does not know the errors of the analysis \cite{Milton2}.
In our calculation of the regularized energy, we start with a determination of 
$\la\tilde{\theta}^{00}\ra$, where the only source of substantial errors is 
the fitting procedure to the Laurent type series. 
However, these errors can be estimated and made smaller in 
more extensive numerical calculations.

\subsection{The regularized canonical vacuum energy}
Following Bender and Hays \cite{Bender}, 
we regularize the canonical 
energy $E$ by integrating $\la\tilde{\theta^{00}}(r')\ra$ 
only to an upper limit of  
\dmb
r'_{\text{max}}=(1-\ep'), \ \ \ep':=\frac{\ep}{R}\ ,
\dme
where $\ep$ denotes the distance from the boundary. Using Eq.\ (\ref{Las}), 
we obtain
\eab
\label{ene}
E({\ep'})&=&\frac{1}{R}\int_0^{1-\ep'}dr^\prime\ 
{r^\prime}^{\ 2}\la\tilde{\theta}^{00}(r')\ra\nonumber\\ 
&=&\frac{1}{R}\ \bigg[\tilde{c}_{-3}\ {\ep^\prime}^{\ -3}+
\tilde{c}_{-2}\ {\ep^\prime}^{\ -2}+\tilde{c}_{-1}\ {\ep^\prime}^{\ -1}
\nonumber\\
&&\phantom{\frac{1}{R}\ \bigg[}{}+
\tilde{c}_{\text{log}}\ \mbox{log}\ \ep^\prime+
\tilde{c}_0+{\cal O}(\ep')\bigg]\ .
\eae
The coefficients $\{c\}$ of Eq.\ (\ref{Las}) and 
the coefficients $\{\tilde{c}\}$ of Eq.\ (\ref{ene}) are related by
\eqb
\label{cneg}
\begin{array}{rcl}
\tilde{c}_{-3}&=& \frac{1}{3}\ c_{-4}, \\ 
 \tilde{c}_{-2}&=& -c_{-4}+\frac{1}{2}\ c_{-3},\\ 
\tilde{c}_{-1}&=& c_{-4}-2\ c_{-3}+c_{-2},\\  
 \tilde{c}_{\text{log}}&=&-c_{-3}+2\ c_{-2}-c_{-1},
\end{array}
\eqe
for the divergent terms,
whereas the coefficients of the positive powers of $\ep^\prime$ 
depend also on the truncation number $N$. Since we are 
only interested in the limit $\ep^\prime \to 0$, the only 
substantial coefficient of $\{\tilde c_j\,|\,j\ge 0\}$  is $\tilde c_0$, given by
\eqb
\tilde{c}_0=-\frac{1}{3}\ c_{-4}+
\frac{3}{2}\ c_{-3}-\frac{3}{2}\ c_{-1}+
\sum_{i=0}^N\ \frac{1}{i+3}\ c_i \ .
\eqe

For the fermionic case, we may assume $c_{-4}=0$. This is so for fermionic quantum fields 
because the canonical 
energy-momentum tensor coincides with the improved tensor, and 
hence it is traceless \cite{Deutsch}. 

Table \ref{T1} contains a list of the coefficients $\{\tilde{c}\}$ 
for the various fields and boundary conditions, where the errors 
are determined from the errors of the coefficients $\{c\}$ 
according to the rules of error propagation. Note that the leading divergences for 
the scalar Dirichlet and Neumann fields exhibit a behavior analogous to that of the
half-space problem, i.e., they are equal in magnitude and carry opposite signs.

\section{The fermionic bag constant}
The bag constant $B$ is 
 introduced into the Lagrangian to achieve conservation of a 
Poincar\'{e} generator, i.e.\ the energy \cite{Chodos74a}. In bag model 
calculations $B$ is determined using the minimization condition 
\dmb
\frac{d}{d R}\ E(R)=0\ ,
\dme
where $E(R)$ denotes the total energy of the bag. Milton \cite{Milton} 
suggests that one should calculate the bag constant from first principles, 
by interpreting it as a vacuum expectation value. The nonlinear boundary
condition
of the model then serves as a definition for $B$. 
Here we will carry out the calculation of this 
vacuum expectation value for a massless fermion field, 
subject to the boundary conditions
of the M.I.T.\  bag model on a static 
sphere with radius $R$. 
The expression for $B_{\text {MIT,q}}$ reads
\eqb
\label{B}
B_{\text {MIT,q}}:=
\left.-\frac{1}{2}\la\pad_r\ (\bar{\psi}\psi)\ra\right|_{r=R}
\eqe
and implies that the differential operator
\eqb
\label{OB}
-\lim_{y\to x}\ \frac{1}{2}\ \text{Tr}\ \left(\pad_{r_x}+\pad_{r_y}\right)
\eqe
should  be applied to the difference of the cavity and the free-space 
propagator. 
We expect $B_{\text {MIT,q}}$ to be infinite. 
In order to 
regularize it, we compute it 
at some interior point a distance $\ep$ away from the boundary,
rather than
at some point on the boundary, as Eq.\ (\ref{B}) demands.
 For the angular   
integrated version $\tilde{B}_{\text {MIT,q}}(\ep')$ 
of $B_{\text {MIT,q}}(\ep')$, we arrive at
\eab
\label{Be}
\lefteqn{
\tilde{B}_{\text {MIT,q}}(\ep'):=4\pi\ B_{\text {MIT,q}}(\ep')}\nonumber \\ 
&=&-\frac{1}{\pi^{1/2}} 
\int_0^{\infty}\!\!\! dz \ \frac{1}{z^{1/2}}\ 
\sum_{\kappa}(2J+1) \sum_{{n>0}}\ {\cal N}^2_{n,\kappa}\ \ep_{n,\kappa}^2
\e^{-z\ep_{n,\kappa}^2}\
\nonumber\\ 
& &{}\times \left\{
\frac{j_{l}(|\ep_{n,\kappa}|r)}
{2l+1}\ \left(l\ j_{l-1}(|\ep_{n,\kappa}|r)-
(l+1)\ j_{l+1}(|\ep_{n,\kappa}|r)\right)\right.\nonumber\\ 
& & \left.
{}-\frac{j_{\bar{l}}(|\ep_{n,\kappa}|r)}
{2\bar{l}+1}\ \left(\bar{l}\,j_{\bar{l}-1}
(|\ep_{n,\kappa}|r)-
(\bar{l}+1)
j_{\bar{l}+1}(|\ep_{n,\kappa}|r)\right)\right\} 
 , \nonumber\\ 
\eae
with
\dmb
\ep'=\frac{\ep}{R}\ ,\ \ r':=1-\ep'\ . 
\dme
There is no free-space part in Eq.\ (\ref{Be}) since 
$\text{Tr}\ \gamma^\mu\equiv 0$ for $\mu=0,\dots,3$.

The result of the calculation of the 
coefficients $\{\tilde{c}\}$ in the expansion of the regularized bag constant 
energy $E^B_{\text {MIT,q}}(\ep')$ with 
\eab
\label{Bene}
E^B_{\text {MIT,q}}(\ep')&:=&\frac{R^3}{3}(1-\ep')^3\times\tilde{B}_{\text {MIT,q}}(\ep')\nonumber\\ 
&=&\frac{1}{R}\ \left[\tilde{c}_{-4}\ {\ep^\prime}^{\ -4}+
\tilde{c}_{-3}\ {\ep^\prime}^{\ -3}+\tilde{c}_{-2}\ {\ep^\prime}^{\ -2}
\right.\nonumber\\
&&\phantom{\frac{1}{R}\ \Bigg[} +\left.
\tilde{c}_{-1}\ {\ep^\prime}^{\ -1}+
\tilde{c}_0+{\cal O}(\ep')\right]\ 
\eae
is displayed in Table \ref{T3}. 
Here we also list the set of coefficients in the expansion of the total  
vacuum energy $E^{\text {tot}}_{\text {MIT,q}}(\ep')$ given by   
\eab
E^{\text{tot}}_{\text {MIT,q}}(\ep')&:=&\int_{0}^{1-\ep'}
\!\!\!\!\!\! dr'\, (r')^2 \left\{
\la\tilde{\theta}^{00}(r')\ra_{\text {MIT,q}}-
\tilde{B}_{\text {MIT,q}}(\ep')\right\}\nonumber\\ 
&=&E_{\text {MIT,q}}(\ep')-E^B_{\text {MIT,q}}(\ep')\ .
\eae

\section{Summary and Discussion }
The main purpose of this paper was to investigate the effect of a static and
spherical boundary on the canonical vacuum energy
densities of otherwise free 
massless Klein-Gordon and Dirac fields. 
Thereby the linear boundary conditions of the 
M.I.T.\ bag model \cite{Chodos74a} have been
used. A Green's function method, that is based on the eigenmode representation 
of the propagator 
and the Schwinger parametrization of the Euclidean momentum squared denominator,
has been used to obtain numerical results for the vacuum energy densities. 
For the fermionic field, a calculation 
of the regularized bag constant based on its definition 
via the  nonlinear boundary condition was carried out. 
Numerical results for the densities have been fitted to Laurent type series in 
powers of the 
distance to the boundary. 
The expressions obtained in this
manner could be integrated within a regularization volume
to yield an expansion of the energies in terms of the regularization 
parameter. Numerical errors of the coefficients in these expansions 
turn out to be less than 1\% for the leading singularity,
and up to 100\% for the weakest divergences.  
In general, the finite part $\tilde{c}_0$ could be determined to about 50\% 
accuracy, using this method and the stated computational effort.

We can compare our results  directly with those in the literature. 
Bender and Hays \cite{Bender} have calculated the 
leading divergence of the canonical part of the vacuum energy 
using a Green's function method. They introduced their regularization 
in the same fashion as we do, but they
relied on analytical formulae for the asymptotic expansion of Bessel functions. 
For example in the fermionic and the Dirichlet scalar case, 
the comparison is as follows:
\eqb
\begin{array}{r@{\ \ }rcl}
\mbox{Bender \& Hays}: & \tilde{c}^{\text {MIT,q}}_{-2}&=& -\frac{1}{120\pi}\approx-0.0027\ ,
\\
&  \tilde{c}^{\text {D}}_{-3}&=& -\frac{1}{24\pi}\approx-0.0133\\ 
\mbox{this work}:&  \tilde{c}^{\text {MIT,q}}_{-2}&=& -0.01058\pm0.45\%\ ,
\\ &  \tilde{c}^{\text {D}}_{-3}&=& \phantom{-}0.01324\pm0.1\%
\end{array}
\eqe
There are, of course, disagreements, 
but the sign error of Bender and Hays' result for the scalar case
has already indirectly been pointed out by Milton \cite{Milton2}. In his work 
he mentions an overall sign error in the vector field mode sum of Ref.\ \cite{Bender}. 
Since the mode sum of the transverse electric vector field
 is up to the $l=0$ contribution 
equal to that of the scalar Dirichlet case (compare Eqs.\ (2.27) and (2.15) in 
Ref.\ \cite{Bender}), 
an overall sign error implies a sign error in 
the result of the scalar Dirichlet case. The factor of four 
difference in the fermionic case might be due to an omission of the trace of
$\gamma_0^2$ as
required by Eq.\ (\ref{OD}). 

Olaussen and Ravndal \cite{Rav} used a Green's function 
method to calculate the {\em electromagnetic} canonical vacuum energy density 
for a spherical bag, and we are thus not able to compare their results with 
ours. 

There are two common types of regularization procedures.
For example Milton \cite{Milton,Milton2} calculated 
the canonical fermionic and vector field vacuum energy using a temporal regularization, whereas 
Bender and Hays \cite{Bender} worked with the same volume regularization as we did.
 Temporal regularization derives from the idea of point 
splitting as a means of cutting off ultraviolett 
contributions to the energy density at each point in the cavity. 
On the other hand, the  volume regularization used above
avoids the integration of {\em boundary} divergences of the energy density. 
Therefore a strict comparison of results obtained 
in those two different regularization schemes should 
not be made. One would, however, 
expect the same divergence structure with different coefficients. 
An order of magnitude check of Milton's result in comparison with our 
result for $\tilde{c}_{-2}$ 
in the fermionic case \cite{Milton} reveals less than a factor 
of 10 difference. The large deviation in the finite parts (a factor of about $10^2$) 
can very well be a consequence of the use 
of asymptotic expansions in Ref.\ \cite{Milton}. In fact, as Milton himself points 
out in the conclusions of Ref. \cite{Milton2}, 
even existing logarithmic divergences are then not recognized.

The small error of the leading infinite terms allows
us to compare the results for the canonical and the bag constant 
part in the fermionic case. A 
cancellation 
of the infinities between these two contributions to the 
total energy does {\em not} occur, since there is a nonvanishing
quartic divergence in the bag constant 
contribution, whereas the leading infinite term in the canonical part
is quadratic (see Table \ref{T3}).
 Moreover, the coefficient $\tilde{c}_0$ in $E^B_{\text {MIT,q}}$ is 
quite large. 

One may speculate that, in a gauge theory 
calculated perturbatively,
infinities arising from the canonical and the bag constant part of the
energy 
 should cancel each other,
leaving a meaningful  finite part. 
We plan to investigate this matter in the near future. 
If there is still no hint for a 
cancellation, one would argue that the introduction of a {\em constant} 
bag energy density $B$ is a too naive device
to achieve the conservation of the Poincar\'{e} generator, i.e.\ the energy. 
Perhaps, a locally conserved energy-momentum tensor 
(if at all definable on physical grounds) 
would reveal finite vacuum energies.  
On the other hand, using  the canonical energy-momentum tensor together with 
soft boundaries, i.e.\ practically confining potentials 
(as for example a harmonic oscillator potential), 
could possibly bypass the global vacuum infinities. The drawbacks of this 
method are the enormous
numerical effort and the necessity of a cutoff parameter to ensure confinement. 
Moreover, it is quite unsatisfactory, that
this additional parameter would have to be determined experimentally. 
Along the same lines, there have been suggestions to introduce 
phenomenological parameters 
in the expression for the bag energy. These parameters could absorb the 
divergences. However, they should be determined
experimentally \cite{Bordag,Milton}, which is again unsatisfactory.

A possible extension of the work done here would 
be a more extensive numerical calculation 
(determination of the energy density at points closer to the boundary) 
to achieve higher precision for the coefficients of the subleading divergences and 
the finite part.

\begin{acknowledgement}
One of us (R.H.) would like to express his gratitude for the warm 
hospitality extended to him at the Physics Department of 
the University of Cape Town. Financial contributions from the Foundation for
Fundamental Research (FFR) and the Gra\-du\-ier\-ten\-kolleg 
``Struktur und Wechselwirkung von Hadronen und Kernen'' 
are gratefully acknowledged. 
\end{acknowledgement}

\appendix
\section{Cavity modes}

Here, the eigenmodes 
and propagators for massless scalar and Dirac fields in a static 
spherical cavity of radius $R$ are given \cite{Buser,OConnor,Viollier83}.

\subsection{Massless scalar fields}
\label{appscalar}

The cavity modes of the massless scalar fields are given as
\be
\phi_{n,l,\mu}(\vec x) = 
\frac{{\cal N}_{n,l}^{\text{D,N}}}{R^{3/2}}\, i\, 
j_l(\varepsilon_{n,l} r) Y_{l\mu}(\hat r) \ .
\ee
A mode representation of the propagator reads as
\eab
\Delta(x,x')&=&
i\sum_{n,l,\mu}
\phi_{n,l,\mu}(\vec{x})\phi^\ast_{n,l,\mu}(\vec{x}')\nonumber\\
&&\qquad{}\times\int\!\frac{d\omega}{2\pi}\frac{e^{-i\omega(x_0-x_0')}}
                                {\omega^2-(\varepsilon_{n,l})^2+i0}
\, .				
\eae
The energy eigenvalues $\varepsilon_{n,l}$ 
depend, of course, on the chosen boundary condition. 
In the Dirichlet case, we have the eigenvalue equation 
\be
j_l(\varepsilon_{n,l} R) = 0\ ,
\ee
and in the case of Neumann boundary conditions 
the eigenvalue equation is
\be
l j_l(\varepsilon_{n,l} R) - \varepsilon_{n,l} R \, j_{l+1}(\varepsilon_{n,l} R) = 0 \ .
\ee
The normalization 
constants ${\cal N}^{\text D}_{n,l}$ and ${\cal N}^{\text N}_{n,l}$ are given as
\begin{eqnarray}
\label{normn}
{\cal N}_{m\Sigma}^{\rm D,N} &=& \sqrt{\frac{2}{(\varepsilon_m^\Sigma R)^2 - J(J+1)}}
    \left|\frac{\varepsilon_m^\Sigma R}{j_J(\varepsilon_m^\Sigma R)}\right|,\\    
&&\quad\Sigma=\cal S, \cal L, \cal M  \nonumber\\ 
\label{norme}
{\cal N}_{m\cal E}^{\rm D,N} &=& \frac{\sqrt{2}}{\left|j_{J+1}(\varepsilon_m^{\cal E} R)\right|}
\ .
\end{eqnarray}

\subsection{Massless fermion fields}
\label{appfermi}
The cavity modes for massless fermion fields 
satisfying the linear boundary condition of the M.I.T.\ bag model
are given by the Dirac spinors
\be q_{n,\kappa,\mu}({\vec x}) = \left({g_{n,\kappa}({r})  
\chi^{\mu}_{\kappa}({ \hat r})}               
 \atop {i  
f_{n,\kappa}({r}) \chi^{\mu}_{-\kappa}({ \hat r})}\right),   
\label{solution} \ee 
where $\chi^{\mu}_{\kappa}({ \hat  r})$ is the usual two-component spherical spinor. 
Here $n$, $\kappa$, and $\mu$ denote the radial, Dirac, and magnetic quantum  
numbers  respectively, and the radial functions $g_{n,\kappa}(r)$ and $f_{n,\kappa}(r)$ 
are given in terms of the spherical 
Bessel functions $j_l$ by 
\begin{eqnarray} 
g_{n,\kappa}(r) & = & \frac{{\cal N}_{n,\kappa}}{R^{3/2}}\,j_l(p_{n,\kappa} r) \label{defgn}\\ 
f_{n,\kappa}(r) & = & \frac{{\cal N}_{n,\kappa}\, \sgn(n)\, 
\sgn(\kappa)}{R^{3/2}}j_{\bar l}(p_{n,\kappa}  
r)\ .
\label{deffn}
\end{eqnarray}  
The discrete momenta $p_{n,\kappa}$ in Eqs.\ (\ref{defgn})
and (\ref{deffn}) are determined 
by the linear boundary condition
\be (i \vec\gamma \cdot \hat r
+1)\,q_{n,\kappa,\mu}({\vec r})|_{r=R}=0 \label{boundary}\ee 
which leads to the eigenvalue equation 
\be j_l(x_{n,\kappa})+\sgn(n)\, \sgn(\kappa)\, j_{\bar l}(x_{n,\kappa})=0\ . 
\label{eigenquark} \ee 
The normalization constant ${\cal N}_{n,\kappa}$ 
in Eqs.\ (\ref{defgn}) and (\ref{deffn}) 
is given by 
\be {\cal N}_{n,\kappa} = (2 \omega_n (\omega_n + \kappa))^{-1/2}  
 \left|\frac{x_n}{j_l(x_n)}\right|\ . \ee 
Here we  have introduced the dimensionless energy and momentum  parameters 
\begin{eqnarray}
x_{n,\kappa} &=& p_{n,\kappa} R,\\
\omega_{n,\kappa} &=& \sgn(n)\, x_{n,\kappa}\ .
\end{eqnarray}
The cavity propagator for massless fermions can be represented in terms of these
cavity modes as
\be
S\left(x,x^\prime\right)=
i
\sum_{\kappa\nu\mu}q_{n,\kappa,\mu}(\vec{x})\bar q_{n,\kappa,\mu}(\vec{x}^\prime)
 \int\frac{d\omega}{2\pi}\frac{e^{-i\omega(x_0-x_0')}}{\omega-\varepsilon_n\pm
i0}\ ,
\ee
where the usual Feynman prescription for the poles is used.

\newpage
\begin{table*}
\caption{The coefficients $\{\tilde{c}\}$ 
for the divergent 
and the finite parts of
the canonical vacuum energy $E(\ep')$
 for massless scalar Dirichlet (D) and  
 Neumann (N), and Dirac (MIT,q) fields.\label{T1}} 
\begin{center}
\begin{tabular}{cr@{\ }c@{\ }lr@{\ }c@{\ }lr@{\ }c@{\ }lr@{\ }c@{\ }lr@{\ }c@{\ }l}
\hline
    &
\multicolumn{3}{c}{$\tilde{c}_{-3}$}
& 
\multicolumn{3}{c}{$\tilde{c}_{-2}$}
& 
\multicolumn{3}{c}{$\tilde{c}_{-1}$} 
& 
\multicolumn{3}{c}{$\tilde{c}_{\text{log}}$} 
& 
\multicolumn{3}{c}{$\tilde{c}_0$}
\\ 
\hline 
 D\ \ \      
& $0.0132435$&$\pm$&$0.1\%$  
& $-0.0232$&$\pm$&$10\%$ 
& $-0.0065$&$\pm$&$40\%$ 
& $-0.0052$&$\pm$&$10\%$ 
& $-10.2$&$\pm$&$50\%$ \\ 
 N\ \ \      
& $-0.01318$&$\pm$&$0.5\%$  
& $-0.0012$&$\pm$&$60\%$
& $-0.085$&$\pm$&$40\%$ 
& $0.64$&$\pm$&$15\%$ 
& $45$&$\pm$&$60\%$ \\ 
 MIT,q\ \    
& &0& 
& $-0.01058$&$\pm$&$0.45\% $
& $-0.35$&$\pm$&$100\%$ 
& $-0.73$&$\pm$&$100\% $
& $0.16$&$\pm$&$20\%$ \\  
\hline
\end{tabular}
\end{center}
\end{table*}

\begin{table*}
\caption{The coefficients $\{\tilde{c}\}$ 
for the divergent 
and the finite part of $E^B_{\text {MIT,q}}$ 
and the total vacuum energy $E^{\text {tot}}_{\text {MIT,q}}(\ep')$.\label{T3}}
\begin{center}
\begin{tabular}{crrrrrr}
\hline
& 
\multicolumn{1}{c}{$\tilde{c}_{-4}$}  &
\multicolumn{1}{c}{$\tilde{c}_{-3}$}  &  
\multicolumn{1}{c}{$\tilde{c}_{-2}$}  &  
\multicolumn{1}{c}{$\tilde{c}_{-1}$}  & 
\multicolumn{1}{c}{$\tilde{c}_{\text{log}}$} &  
\multicolumn{1}{c}{$\tilde{c}_{0}$} \\
\hline 
 $E^B_{\text {MIT,q}}$ & --0.1593$\pm$0.2\% & 0.375$\pm$1\% & --0.119$\pm$5\%& 0.11$\pm$20\%& 
 \multicolumn{1}{c}{0} 
 & --7.2$\times10^5\pm$50\% \\  
 $E^{\text {tot}}_{\text {MIT,q}}$ & 0.1593$\pm$0.2\% &--0.375$\pm$1\% & 
0.108$\pm$6\% &--0.46$\pm$85\% & --0.73$\pm$100\% & 7.2$\times10^5\pm$50\%  
\\
\hline
\end{tabular}
\end{center}
\end{table*}

\end{document}